\documentclass[12pt]{article}

\begin{document}
\title{Noether symmetries for two-dimensional charged particle motion}
\author{F.~Haas and J.~Goedert \\
Centro de Ci\^encias Exatas e Tecnol\'ogicas, UNISINOS\\
Av. Unisinos, 950\\
93022--000 S\~ao Leopoldo, RS - Brazil}
\date{\strut}
\maketitle

\begin{abstract}
We find the Noether point symmetries for non--relativistic
two-dimensional charged particle motion. These symmetries are
composed of a quasi--invariance transformation, a time--dependent
rotation and a time--dependent spatial translation. The associated
electromagnetic field satisfy a system of first--order linear
partial differential equations. This system is solved exactly,
yielding three classes of electromagnetic fields compatible with
Noether point symmetries. The corresponding Noether invariants are
derived and interpreted.

\end{abstract}

\section{Introduction}

There exist several methods for the derivation of exact
invariants (constants of motion or first integrals) for dynamical
systems \cite{Kaushal}. Among these methods, special attention has
been focused on the use of Noether's theorem \cite{Noether,
Sarlet} due to its physical appeal. Noether's theorem stablishes a link
between the continuous symmetries of the action functional
associated to a dynamical system and its conservation laws. The
classical examples of Noether's theorem are the conservation of
energy associated to time translation invariance, the conservation
of linear momentum associated to space translation invariance and
the conservation of angular momentum associated to invariance
under rotations.

In the present work, we investigate the Noether point symmetries
for two-dimensional non-relativistic charged particle motion. This
class of systems is described by Lagrangians of the form
\begin{equation}
\label{eq11}
L = \frac{1}{2}(\dot{x}^2 + \dot{y}^2) + A_{1}(x,y,t)\dot{x}
+ A_{2}(x,y,t)\dot{y} - V(x,y,t) \,,
\end{equation}
where, in appropriated units, ${\bf A} = (A_{1}(x,y,t),
A_{2}(x,y,t), 0)$ is the vector potential and $V(x,y,t)$ is the
scalar potential. The results presented here can be useful in the
derivation of exact time--dependent solutions of the
Vlasov--Maxwell equations in plasma physics \cite{Schrauner}.

In the course of the search for Noether conserved quantities, we
find that the general form of Noether point symmetries for
Lagrangians of the class (\ref{eq11}) comprises only
time-dependent rescalings, rotations and spatial translations.
Moreover, the associated electromagnetic fields are constrained by
a pair of linear, first-order partial differential equations. The
form of this pair of equations depends on the symmetry considered.
Nevertheless, for the solution of these equations,
we may always resort to the use of canonical group coordinates,
defined in section 3. Canonical group coordinates
are, in fact, a valuable tool for systematic determination of
the general solution of the basic system of equations satisfied by
the electromagnetic fields compatible with Noether's point
symmetries. We find three categories of such electromagnetic
fields, one corresponding to an energy-like constant of motion,
one to an angular momentum-like constant of motion and, finally,
one with an associated linear momentum-like first integral. These
conserved quantities are gauge independent, a fact that is
explicitly shown in all cases treated here. This shows that the
choice of gauge has no influence in the determination of Noether's
conserved quantities, in contrary to what is sometimes claimed
\cite{Neuenschwander}.

The paper is organized as follows. In section 2, we present
Noether's theorem for the case of point symmetries. The condition
for invariance of the action functional with a Lagrangian of type
(\ref{eq11}) yields both the form of the Noether symmetries and
the equations to be satisfied by the electromagnetic field. These
equations, listed in (\ref{eq27}--\ref{eq29}) below, are the basic
equations to be satisfied by the electromagnetic field in
order to exist the corresponding Noether point symmetries. The
system (\ref{eq27}--\ref{eq29}) is solved by use of canonical
group variables for each type of Noether symmetry. These canonical
group variables are shown in section 3. The corresponding
solutions to system (\ref{eq27}--\ref{eq29}) is given in section
4. In order that the solutions so obtained qualify as true
electromagnetic fields, the Maxwell equations are imposed for
consistency. In section 5, the Noether invariant associated to
each symmetry is showed and interpreted. Finally, section 6 is
devoted to the conclusions.

\section{Noether point symmetries}

Noether's theorem provides a link between continuous symmetries for
the action functional
\begin{equation}
\label{eq0}
S = \int\,L\,dt
\end{equation}
and conserved quantities. In its original and more powerful
formulation \cite{Noether, Sarlet}, Noether's theorem considers
dynamic symmetries involving velocities and higher derivatives.
Here, mainly for simplifying reasons, we  restrict
considerations to point transformations. In more formal terms, we
 consider infinitesimal mappings of the form
\begin{eqnarray}
\label{eq4}
\bar{x} &=& x + \varepsilon\eta_{1}(x,y,t) \,,\\
\label{eq5}
\bar{y} &=& y + \varepsilon\eta_{2}(x,y,t) \,,\\
\label{eq6}
\bar{t} &=& t + \varepsilon\tau(x,y,t) \,,
\end{eqnarray}
where $\varepsilon$ is an infinitesimal parameter. It is useful to
define the generator of the group of symmetries associated to
(\ref{eq4}--\ref{eq6}) as
\begin{equation}
G = \tau\frac{\partial}{\partial t} + \eta_{1}\frac{\partial}{\partial x}
+ \eta_{2}\frac{\partial}{\partial y} \,.
\end{equation}
The generator $G$  appear frequently in what follows
and it is useful in the definition of canonical group
coordinates, which plays a central role in the systematic
determination of the electromagnetic fields associated to
the symmetries.

The condition for Noether symmetry reads \cite{Sarlet}
\begin{equation}
\label{eq7} \tau\,L_t + {\bf\eta}\cdot\,L_{{\bf q}} + (\dot{\bf\eta} -
\dot\tau\dot{\bf q})\cdot\,L_{\dot{\bf q}} + \dot\tau\,L = \dot F \,,
\end{equation}
where subscripts denote partial derivatives. In equation (\ref{eq7})
and in what follows,
\begin{eqnarray}
\label{eq8} {\bf\eta} &=& (\eta_{1}, \eta_{2}, 0) \,,\\ \label{eq9} {\bf q} &=& (x, y,
0) \,,\\ \label{eq10} F &=& F(x,y,t) \,.
\end{eqnarray}
The condition for
Noether symmetry ensures the invariance of the action functional
under the infinitesimal map
(\ref{eq4}--\ref{eq6}), up to the addition of an irrelevant numerical
constant. Now, as a consequence, the Euler-Lagrange equations become
formally invariant under the map (\ref{eq4}--\ref{eq6}) and
the associated Noether invariant reads
\begin{equation}
\label{eqq8} I = \tau\,(\dot{\bf q}\cdot\,L_{\dot{\bf q}} - L) -
{\bf\eta}\cdot\,L_{\dot{\bf q}} + F \,,
\end{equation}
which, in the present case, has the form
\begin{equation}
\label{eqq9} I = \tau\left(\frac{1}{2}\dot{\bf q}^2 + V\right) -
{\bf\eta}\cdot(\dot{\bf q} + {\bf A}) + F \,.
\end{equation}
Clearly, for point symmetries, the Noether invariant is at most
quadratic in the velocity. This reminds of the work of Lewis
\cite{Lewis}, where he finds quadratic constants of motion for the
three--dimensional non--relativistic motion of a charged particle
in an external electromagnetic field. The work of Lewis, however,
did not obtain the whole variety of solutions. Some of the Noether
invariants derived in section 4 of this paper do not fit into the
framework of reference \cite{Lewis}.

In what follows, we find all the Noether point symmetries
associated to Lagrangian (\ref{eq11}). The Noether point
symmetries are identified by imposing the symmetry condition
(\ref{eq7}) and the Noether invariants are constructed by use
of equation (\ref{eqq9}).

We begin by identifying the Noether point symmetries associated
with the Lagrangian (\ref{eq1}). Inserting $L$ in Noether's
symmetry condition (\ref{eq7}), we arrive at a polynomial in the
velocity components. The coefficients of all monomials of the form
${\dot x}^{m}{\dot y}^n$ must be identically zero. For instance,
the coefficients of $\dot{x}^3$ and $\dot{y}^3$ yield
\begin{eqnarray}
\label{eq12}
\dot{x}^3 \rightarrow \tau_x &=& 0 \,,\\
\label{eq13}
\dot{y}^3 \rightarrow \tau_y &=& 0 \,.
\end{eqnarray}
The solution for (\ref{eq12}--\ref{eq13}) is
\begin{equation}
\label{eq14}
\tau = \rho^{2}(t) \,,
\end{equation}
$\rho(t)$ an arbitrary function of time. Equation (\ref{eq14})
is taken into account in the continuation.

The coefficients of $\dot{x}^{2}\dot{y}$ and $\dot{x}\dot{y}^2$ give no new
information, as they are identically zero. The coefficients of the
quadratic components of velocity yield
\begin{eqnarray}
\label{eq15}
\dot{x}^2 \rightarrow \eta_{1x} &=& \rho\dot\rho \,,\\
\label{eq16}
\dot{y}^2 \rightarrow \eta_{2y} &=& \rho\dot\rho \,,\\
\label{eq17}
\dot{x}\dot{y} \rightarrow \eta_{1y} + \eta_{2x} &=& 0 \,.
\end{eqnarray}
The general solution of (\ref{eq17}) is
\begin{equation}
\label{eq18}
\eta_1 = \Gamma_x \quad , \quad \eta_2 = - \Gamma_y \quad ,
\end{equation}
where $\Gamma = \Gamma(x,y,t)$ is an arbitrary function. Inserting
eq. (\ref{eq18}) into eqs. (\ref{eq15}--\ref{eq16}), yield
\begin{equation}
\label{eq19}
\Gamma_{xx} = - \Gamma_{yy} = \rho\dot\rho \,.
\end{equation}
The solution to equation (\ref{eq19}) is a quadratic function of
the spatial coordinates,
\begin{equation}
\label{eq20} \Gamma = \frac{\rho\dot\rho}{2}(x^2 - y^2) -
\Omega(t)\,xy + a_{1}(t)\,x - a_{2}(t)\,y + \Gamma_{0}(t) \,,
\end{equation}
for arbitrary functions of time $\Omega(t), \,a_{1}(t), \,a_{2}(t)$
and $\Gamma_{0}(t)$.

By inserting the solution (\ref{eq20}) into eq. (\ref{eq18}), we
obtain
\begin{eqnarray}
\label{eq21} \eta_1 &=& \rho\dot\rho\,x - \Omega(t)\,x + a_{1}(t)
\,,\\ \label{eq22} \eta_2 &=& \rho\dot\rho\,y + \Omega(t)\,y +
a_{2}(t) \,.
\end{eqnarray}

Let us summarize our first results: up to terms quadratic in the
velocity, the generator of Noether point symmetries has the
general form
\begin{equation}
\label{eq23} G = G_Q + G_R + G_T \,,
\end{equation}
where
\begin{equation}
\label{eqq23}
G_Q = \rho^{2}(t)\frac{\partial}{\partial t} +
\rho\dot\rho\,(x\frac{\partial}{\partial x} + y\frac{\partial}{\partial y})
\quad ,
\end{equation}
is associated to quasi--invariance transformations \cite{Munier},
\begin{equation}
\label{eqq24}
G_R =  \Omega(t)\,(x\frac{\partial}{\partial y} - y\frac{\partial}{\partial
x})
\end{equation}
is associated to time--dependent rotations and
\begin{equation}
\label{eqq25}
G_T = a_{1}(t)\frac{\partial}{\partial x} +
      a_{2}(t)\frac{\partial}{\partial y}
\end{equation}
corresponds to time--dependent spatial translations. It is also
important to notice that, so far, no constraint has been imposed
on the electromagnetic fields.

The Noether symmetry condition, however, has not yet been fully taken
into account. We still must consider the terms linear in
the velocity components, yielding
\begin{eqnarray}
\label{eq24} \dot x \rightarrow F_x &=& G\,A_1 + \rho\dot\rho\,A_1
+ \Omega\,A_2 + (\rho\ddot\rho + \dot\rho^2)\,x - \dot\Omega\,y +
\dot{a}_1 \,,\\ \label{eq25} \dot y \rightarrow F_y &=& G\,A_2 +
\rho\dot\rho\,A_2 - \Omega\,A_1 + (\rho\ddot\rho + \dot\rho^2)\,y
+ \dot\Omega\,x + \dot{a}_2 \,.
\end{eqnarray}
Moreover, the term independent of velocities gives
\begin{eqnarray}
F_t &=& - G\,V - 2\rho\dot\rho\,V + \left((\rho\ddot\rho +
\dot\rho^2)\,x - \dot\Omega\,y + \dot{a}_1\right)\,A_1 + \nonumber
\\
\label{eq26} &+& \left((\rho\ddot\rho + \dot\rho^2)\,y +
\dot\Omega\,x + \dot{a}_2\right)\,A_2 \,.
\end{eqnarray}

The system (\ref{eq24}--\ref{eq26})  has a solution
$F(x,y,t)$ if and only if the integrability conditions $F_{xy} =
F_{yx}$, $F_{xt} = F_{tx}$ and $F_{yt} = F_{ty}$ are fulfilled.
These requirements give equations not for the electromagnetic
potentials, but directly for the electric field ${\bf E} =
(E_{1}(x,y,t), E_{2}(x,y,t), 0)$ and the magnetic field ${\bf B} =
(0, 0, B(x,y,t))$, defined by
\begin{eqnarray}
\label{eq1} E_{1}(x,y,t) &=& -  V_x - A_{1t} \,,\\ \label{eq2}
E_{2}(x,y,t) &=& -  V_y - A_{2t} \,,\\ \label{eq3} B(x,y,t) &=&
A_{2x} - A_{1y} \,.
\end{eqnarray}
In fact, imposing $F_{xy} = F_{yx}$ yields
\begin{equation}
\label{eq27} G\,B =  - 2\rho\dot\rho\,B - 2\dot\Omega \,,
\end{equation}
which involves only the magnetic field. Imposition of $F_{xt} =
F_{tx}$ implies
\begin{eqnarray}
G\,E_1 &=& - 3\rho\dot\rho\,E_1 - \Omega\,E_2 -
\left((\rho\ddot\rho + \dot\rho^2)y + \dot\Omega\,x +
\dot{a}_{2}\right)\,B + \nonumber \\ \label{eq28} &+&
(\rho{\buildrel \cdots\over\rho} + 3\dot\rho\ddot\rho)\,x -
\ddot\Omega\,y + \ddot{a}_1 \,,
\end{eqnarray}
whereas $F_{yt} = F_{ty}$ implies
\begin{eqnarray}
G\,E_2 &=& - 3\rho\dot\rho\,E_2 + \Omega\,E_1 +
\left((\rho\ddot\rho + \dot\rho^2)x - \dot\Omega\,y +
\dot{a}_{1}\right)\,B + \nonumber \\ \label{eq29} &+&
(\rho{\buildrel \cdots\over\rho} + 3\dot\rho\ddot\rho)\,y +
\ddot\Omega\,x + \ddot{a}_2 \,.
\end{eqnarray}

Equations (\ref{eq27}--\ref{eq29}) are the equations to be
satisfied by the electromagnetic fields associated to Noether
point symmetries. They constitute a system of linear, first order,
coupled partial differential equations for $E_1$, $E_2$ and $B$.
For each non-relativistic charged particle motion under an
electromagnetic field satisfying (\ref{eq27}--\ref{eq29}), there
is an associated Noether point symmetry, whose generator is given
by eq. (\ref{eq23}). In the remaining of this work, we are
essentially interested in finding all the solutions of the system
of partial differential equations (\ref{eq27}--\ref{eq29}). In
other words, we are concerned with finding the most general
electromagnetic field for two-dimensional, non-relativistic
charged particle motion endowed with Noether point symmetries. It
is interesting to note that, while Noether's theorem is formulated
in terms of the action functional, which is gauge dependent, the
final conditions for Noether point symmetry involves only the
physical fields and the symmetry generator and not the potentials.
This means that the choice of gauge does not play any role in the
search for Noether point symmetries. Another useful remark is that
$B$ satisfies an equation decoupled from the equations for $E_1$
and $E_2$, whereas the equations for the electric field do depend
on $B$. Thus, we must first solve (\ref{eq27}) for $B$ and
afterwards solve (\ref{eq28}--\ref{eq29}) for the electric field.

The unknown functions are the electromagnetic potentials ${\bf A}$
and $V$. Therefore, for any solution ${\bf E}$, ${\bf B}$ of the
system (\ref{eq27}--\ref{eq29}), we still have to solve equations
(\ref{eq1}--\ref{eq3}) to obtain the electromagnetic potentials.
It turns out that the integrability condition for the system
(\ref{eq1}--\ref{eq3}) are the homogeneous Maxwell equations.
Gauss law for magnetism is immediately satisfied here. So, the
only extra requirement that we must impose is Faraday's law
\begin{equation}
\label{eq30} E_{2x} - E_{1y} + B_t = 0 \,.
\end{equation}
With that last constraint, the solutions ${\bf E}$, ${\bf B}$ for
the basic system (\ref{eq27}--\ref{eq29}) qualifies as a true
electromagnetic field.

To treat the system (\ref{eq27}--\ref{eq29}) and to find its
complete solution, we use canonical group coordinates. These
variables are be introduced in the section that follows.

\section{Canonical group coordinates}

Canonical group coordinates \cite{Bluman} are defined by imposing
that the symmetry transformation behave merely like  time
translation. Denoting new coordinates by
$(\bar{x},\bar{y},\bar{t})$, such condition means that, in
canonical group coordinates,
\begin{equation}
\label{eq31}
G = \frac{\partial}{\partial\bar{t}} \,,
\end{equation}
where $\bar{t}$ is the new time parameter. The equations which must be
satisfied by any set of canonical group coordinates are given by
\begin{equation}
\label{eq32} G\,\bar{x} = 0 \quad , \quad G\,\bar{y} = 0 \quad ,
\quad G\,\bar{t} = 1 \quad .
\end{equation}
This is a set of uncoupled linear partial differential equations,
which, for the generator (\ref{eq23}), can be solved in closed
form by the method of characteristics. We find three classes of
solutions, listed below.

\subsection{The case $\rho \neq 0$}

When $\rho \neq 0$, it is convenient to write
\begin{eqnarray}
\label{eqq33}
a_1 &=& \rho(\rho\dot\alpha_{1} - \dot\rho\alpha_{1}) \,,\\
\label{eqq34}
a_2 &=& \rho(\rho\dot\alpha_{2} - \dot\rho\alpha_{2})
\end{eqnarray}
for suitable functions $\alpha_{1}(t)$ and $\alpha_{2}(t)$,
defined in terms of $a_1$ and $a_2$. Notice that for $\rho = 0$
the transformation (\ref{eqq33}--\ref{eqq34}) is meaningless and
the case is treated separately.

With the redefinition (\ref{eqq33}--\ref{eqq34}), we have the
following canonical group coordinates,
\begin{eqnarray}
\label{eq33} \bar{t} &=& \int^{t}d\mu/\rho^{2}(\mu) \,,\\
\label{eq34} \bar{x} &=& \frac{(x-\alpha_{1})}{\rho}\cos\,T +
\frac{(y-\alpha_{2})}{\rho}\sin\,T + \delta_{1}  \,,\\
\label{eq35} \bar{y} &=& \frac{(y-\alpha_{2})}{\rho}\cos\,T -
\frac{(x-\alpha_{1})}{\rho}\sin\,T + \delta_{2} \,,
\end{eqnarray}
where new functions $T = T(t), \,\delta_1 = \delta_{1}(t)$ and
$\delta_2 = \delta_{2}(t)$
were defined according to
\begin{eqnarray}
\label{eq36}
T(t) &=& \int^{t}d\mu\,\Omega(\mu)/\rho^{2}(\mu) \,,\\
\label{eq37}
\delta_{1}(t) &=& \int^{t}d\mu
\frac{\Omega(\mu)}{\rho^{3}(\mu)}\left(\alpha_{2}(\mu)
\cos\,T(\mu) - \alpha_{1}(\mu)\sin\,T(\mu)\right) \,,\\
\label{eq38}
\delta_{2}(t) &=& - \int^{t}d\mu
\frac{\Omega(\mu)}{\rho^{3}(\mu)}\left(\alpha_{1}(\mu)
\cos\,T(\mu) + \alpha_{2}(\mu)\sin\,T(\mu)\right) \,.
\end{eqnarray}

As a particular case, let us consider the situation where the
symmetry transformation does not contain rotation, that is,
$\Omega = 0$. In that case, the canonical group variables
(\ref{eq33}--\ref{eq35}) take the form
\begin{eqnarray}
\label{eq39}
\bar{t} &=& \int^{t}d\mu/\rho^{2}(\mu) \,,\\
\label{eq40}
\bar{x} &=& \frac{(x-\alpha_{1})}{\rho} \,,\\
\label{eq41}
\bar{y} &=& \frac{(y-\alpha_{2})}{\rho} \,,
\end{eqnarray}
which are relevant in the study of time--dependent integrable
systems \cite{LL}. For $\alpha_1 = \alpha_2 = 0$, the
transformation (\ref{eq39}--\ref{eq41}) is known as a
quasi-invariance transformation \cite{Munier}.

\subsection{The case $\rho = 0$ and $\Omega \neq 0$}

The canonical group variables become, in this case,
\begin{eqnarray}
\label{eq42}
\bar{x} &=& \left((x - \beta_{1})^2 + (y - \beta_{2})^{2}\right)^{1/2}
\,,\\
\label{eq43}
\bar{y} &=& t \,,\\
\label{eq44}
\bar{t} &=& \frac{1}{\Omega}\tan^{-1}
\left(\frac{y - \beta_{2}}{x - \beta_{1}}\right) \,,
\end{eqnarray}
where we have defined
\begin{equation}
\label{eqq44}
\beta_1 = \beta_{1}(t) = - a_{2}/\Omega \quad , \quad
\beta_2 = \beta_{2}(t) = a_{1}/\Omega
\quad .
\end{equation}
The variables $\bar{x}$ and $\bar{t}$ are translated polar
coordinates, the new time parameter plays the role of an
azimuthal angle and $\bar{x}$ plays the role of a radial coordinate.

\subsection{The case $\rho = 0$, $\Omega = 0$ and $a_{2} \neq 0$}

The canonical group variables are now
\begin{eqnarray}
\label{eq45}
\bar{x} &=& x - a_{1}y/a_{2} \,,\\
\label{eq46}
\bar{y} &=& t \,,\\
\label{eq47}
\bar{t} &=& y/a_2 \,.
\end{eqnarray}
We finally mention that the case $\rho = 0$, $\Omega = 0$ and
$a_{1} \neq 0$ is strictly analogous to the last case and deserve
no special consideration.

In the following section, we obtain all the solutions for the
basic system (\ref{eq27}--\ref{eq29}), corresponding to each set
of canonical group variables.

\section{Electromagnetic fields}

\subsection{The case $\rho \neq 0$}

Equation (\ref{eq27}), which involves only the magnetic field
acquires, in canonical group coordinates, the form
\begin{equation}
\label{eq48}
B_{\bar{t}} = - \frac{2\rho'}{\rho}B - \frac{2\Omega'}{\rho^2} \,,
\end{equation}
where the prime denotes total differentiation with respect to $\bar{t}$.
The solution for (\ref{eq48}) is
\begin{equation}
\label{eq49}
B = - \frac{2\Omega}{\rho^2} + \frac{1}{\rho^2}\bar{B}(\bar{x},\bar{y}) \,,
\end{equation}
where $\bar{B}(\bar{x},\bar{y})$ is an arbitrary function of the
indicated arguments. Notice that the resulting magnetic field is
not necessarily homogeneous, since it can depend on the spatial
coordinates through $\bar{x}$ and $\bar{y}$. This is a significant
improvement on earlier results \cite{Bouquet}.

To find the electric field, we must solve the system
(\ref{eq28}--\ref{eq29}), taking the solution (\ref{eq49}) into
account. To solve (\ref{eq28}--\ref{eq29}), it is useful to
consider the quantities $\Sigma_1$ and $\Sigma_2$ defined by
\begin{eqnarray}
\label{eq50}
\Sigma_1 &=& \rho^{3}(E_{1}\cos\,T + E_{2}\sin\,T) \,,\\
\label{eq51}
\Sigma_2 &=& \rho^{3}(E_{2}\cos\,T - E_{1}\sin\,T) \,.
\end{eqnarray}
This represents a rotation plus a rescaling of the electric field.
The form (\ref{eq50}--\ref{eq51}) represents a
circularly polarized wave with time-dependent amplitude. .

In the new variables, the system (\ref{eq28}--\ref{eq29})
decouples and can be cast into the form
\begin{equation}
\label{eq52}
\frac{\partial\Sigma_1}{\partial\bar{t}} =
\frac{\partial\psi_1}{\partial\bar{t}} \quad , \quad
\frac{\partial\Sigma_2}{\partial\bar{t}} =
\frac{\partial\psi_2}{\partial\bar{t}} \quad ,
\end{equation}
where
\begin{eqnarray}
\psi_1 &=& \left(- \frac{\rho'}{\rho}(\bar{y} - \delta_{2}) +
\delta_{2}' - \Omega(\bar{x} - \delta_{1}) + \frac{1}{\rho}
(\alpha_{1}'\sin\,T - \alpha_{2}'\cos\,T)\right)
\bar{B}(\bar{x},\bar{y}) + \nonumber \\ &+&
\left(\frac{\rho''}{\rho} - 2\frac{{\rho'}^2}{\rho^2} +
\Omega^{2}\right)(\bar{x} - \delta_{1}) - \left(\Omega' -
2\frac{\rho'}{\rho}\Omega\right)(\bar{y} - \delta_{2}) + \nonumber
\\ &+& \frac{1}{\rho}\left(\Omega'\alpha_1 - \Omega(\alpha_{1}' +
\frac{\rho'}{\rho}\alpha_1) + \alpha_{2}'' -
2\frac{\rho'}{\rho}\alpha_{2}'
 + \Omega^{2}\alpha_{2}\right)\sin\,T + \\
&+& \frac{1}{\rho}\left(- \Omega'\alpha_2 + \Omega(\alpha_{2}' +
\frac{\rho'}{\rho}\alpha_2) + \alpha_{1}'' - 2\frac{\rho'}{\rho}\alpha_{1}'
 + \Omega^{2}\alpha_{1}\right)\cos\,T \,, \nonumber \\
&\strut& \nonumber \\ \psi_2 &=& \left(+
\frac{\rho'}{\rho}(\bar{x} - \delta_{1}) - \delta_{1}' -
\Omega(\bar{y} - \delta_{2}) + \frac{1}{\rho} (\alpha_{1}'\cos\,T
+ \alpha_{2}'\sin\,T)\right) \bar{B}(\bar{x},\bar{y}) + \nonumber
\\ &+& \left(\frac{\rho''}{\rho} - 2\frac{{\rho'}^2}{\rho^2} +
\Omega^{2}\right)(\bar{y} - \delta_{2}) + \left(\Omega' -
2\frac{\rho'}{\rho}\Omega\right)(\bar{x} - \delta_{1}) + \nonumber
\\ &-& \frac{1}{\rho}\left(- \Omega'\alpha_2 + \Omega(\alpha_{2}'
+ \frac{\rho'}{\rho}\alpha_2) + \alpha_{1}'' -
2\frac{\rho'}{\rho}\alpha_{1}'
 + \Omega^{2}\alpha_{1}\right)\sin\,T + \\
&+& \frac{1}{\rho}\left(+ \Omega'\alpha_1 - \Omega(\alpha_{1}' +
\frac{\rho'}{\rho}\alpha_{1}) + \alpha_{2}'' - 2\frac{\rho'}{\rho}\alpha_{2}'
 + \Omega^{2}\alpha_{2}\right)\cos\,T \,. \nonumber
\end{eqnarray}

The solution to (\ref{eq52}) is
\begin{equation}
\label{eq55}
\Sigma_1 = \psi_1 + \bar{E}_{1}(\bar{x},\bar{y}) \quad , \quad
\Sigma_2 = \psi_2 + \bar{E}_{2}(\bar{x},\bar{y}) \,,
\end{equation}
where, as indicated, $\bar{E}_1$ and $\bar{E}_2$ have no
dependence on $\bar{t}$.

We are interested in the electric field, in physical variables. To
obtain the physical field we use the inverse of the transformation
(\ref{eq50}--\ref{eq51}),
\begin{eqnarray}
\label{eq56}
E_1 &=& \frac{1}{\rho^3}(\Sigma_{1}\cos\,T - \Sigma_{2}\sin\,T) \,,\\
\label{eq57}
E_2 &=& \frac{1}{\rho^3}(\Sigma_{2}\cos\,T + \Sigma_{1}\sin\,T) \,.
\end{eqnarray}
Substituting equations (\ref{eq56}--\ref{eq57}) into the
solution (\ref{eq55}) and transforming back to the original
variables $(x,y,t)$, we obtain the electric fields
\begin{eqnarray}
E_1 &=& \ddot\alpha_1 + \frac{\ddot\rho}{\rho}(x - \alpha_{1}) +
\frac{\Omega^{2}x}{\rho^4} - (\rho\dot\Omega -
2\dot\rho\Omega)\frac{y}{\rho^3} +
\frac{\Omega}{\rho^3}(\rho\dot\alpha_2 - \dot\rho\alpha_{2}) +
\nonumber \\
\label{eq58}
&+& \frac{1}{\rho^3}\left(\bar{E}_{1}(\bar{x},\bar{y})\cos\,T
- \bar{E}_{2}(\bar{x},\bar{y})\sin\,T\right) \\
&-& \frac{1}{\rho^4}\left(\rho\dot\rho(y - \alpha_2) + \rho^{2}\dot\alpha_2
+ \Omega\,x\right)\bar{B}(\bar{x},\bar{y}) \,, \nonumber \\
E_2 &=& \ddot\alpha_2 + \frac{\ddot\rho}{\rho}(y - \alpha_{2}) +
\frac{\Omega^{2}y}{\rho^4} + (\rho\dot\Omega -
2\dot\rho\Omega)\frac{x}{\rho^3} -
\frac{\Omega}{\rho^3}(\rho\dot\alpha_1 - \dot\rho\alpha_{1}) +
\nonumber \\
\label{eq59}
&+& \frac{1}{\rho^3}\left(\bar{E}_{2}(\bar{x},\bar{y})\cos\,T
+ \bar{E}_{1}(\bar{x},\bar{y})\sin\,T\right) +   \\
&+&
\frac{1}{\rho^4}\left(\rho\dot\rho(x - \alpha_1) + \rho^{2}\dot\alpha_1
- \Omega\,y\right)\bar{B}(\bar{x},\bar{y}) \,.  \nonumber
\end{eqnarray}

It still remains to take into consideration Faraday's law, which,
in our case, is equivalent to eq. (\ref{eq30}). After lengthy
calculations using the magnetic field (\ref{eq49}) and the
electric field (\ref{eq58}--\ref{eq59}), we conclude that
Faraday's law imposes only that
\begin{equation}
\label{eq60}
\bar{E}_{2\bar{x}} - \bar{E}_{1\bar{y}} = 0 \,.
\end{equation}
This last constraint is an equation that has general solution
\begin{equation}
\label{eq61}
\bar{E}_1 = - \frac{\partial}{\partial\bar{x}}\bar{V}(\bar{x},\bar{y})
\quad , \quad
\bar{E}_2 = - \frac{\partial}{\partial\bar{y}}\bar{V}(\bar{x},\bar{y})
\quad ,
\end{equation}
where $\bar{V}(\bar{x},\bar{y})$ is an arbitrary function of the
indicated argument.

In conclusion, we have obtained a very general class of electromagnetic
fields yielding Noether point symmetries. The magnetic field is given by
eq. (\ref{eq49}) and the electric field by eqs. (\ref{eq58}--\ref{eq59}),
together with condition (\ref{eq61}). The symmetry transformations
has the generator (\ref{eq23}). The electromagnetic field involves
several arbitrary functions, namely $\rho(t), \alpha_{1}(t),
\alpha_{2}(t), \Omega(t), \bar{B}(\bar{x},\bar{y})$ and
$\bar{V}(\bar{x},\bar{y})$, where $\bar{x}$ and $\bar{y}$ are
defined by eqs. (\ref{eq34}--\ref{eq35}).

Finally, the electric field may be represented in a much more compact way.
Introducing the vectors
\begin{eqnarray}
\label{eq62} {\bf\alpha} &=& (\alpha_{1}, \alpha_{2}, 0) \,,\\
\label{eq64} {\bf\Omega} &=& (0, 0, \Omega) \,,\\ \label{eq66} \bar{\bf E} &=&
(\bar{E}_{1}, \bar{E}_{2}, 0) \,, \\ \label{eqqq66} \bar{\bf B} &=& (0,0,\bar{B}) \,,
\end{eqnarray}
redefining ${\bf\eta}$
\begin{equation}
\label{eq65} {\bf\eta} = \rho\dot\rho({\bf q} - {\bf\alpha}) + \rho^{2}\dot{\bf\alpha}
+ {\bf\Omega}\times{\bf q} \,,
\end{equation}
and using the rotation matrix
\begin{equation}
\label{eqq66}
R(T) = \left(\matrix{\cos\,T & - \sin\,T & 0 \cr
                     \sin\,T & \cos\,T & 0 \cr
                     0 & 0 & 1\cr}\right) \quad ,
\end{equation}
we can represent the electric field by the form
\begin{equation}
\label{eq68} {\bf E} = \frac{1}{\rho^4}(\rho(\rho{\bf\eta}_t - \dot\rho{\bf\eta}) +
{\bf\eta}\times{\bf\Omega}) + \frac{1}{\rho^3}R(T)\bar{\bf E} +
\frac{1}{\rho^4}\bar{\bf B}\times{\bf\eta} \,,
\end{equation}
where $\bar{\bf E}$ is given in terms of a scalar potential
according to (\ref{eq61}).

\subsection{The case $\rho = 0$ and $\Omega \neq 0$}

In the case $\rho = 0$, the symmetry transformation is
composed of a rotation and a spatial translation and has no
rescaling part. The treatment presented in the last subsection is
no longer appropriate, because the limit $\rho = 0$ of the
canonical variables (\ref{eq33}--\ref{eq35}) is singular. We must
now use the canonical group variables (\ref{eq42}--\ref{eq44}).
The steps to be carried out comprise: a) the calculation of $B$
using the basic equation (\ref{eq27}), b) the calculation of $E_1$
and $E_2$ using the basic equations (\ref{eq28}--\ref{eq29}), and
c) the imposition of Faraday's law, which must be obeyed by the
resulting electromagnetic field.

The equation (\ref{eq27}) for $B$, in canonical group variables
(\ref{eq42}--\ref{eq44}), reads
\begin{equation}
\label{eq69} B_{\bar{t}} = - 2\dot\Omega(\bar{y}) \,,
\end{equation}
having the general solution
\begin{equation}
\label{eq70} B = - 2\dot\Omega(\bar{y})\bar{t} + \bar{B}(\bar{x},
\bar{y}) \,,
\end{equation}
where $\bar{B}(\bar{x}, \bar{y})$ is an arbitrary function of the
indicated arguments.

By the definition (\ref{eq44}) of the new time parameter, however,
it is clear that $\bar{t}$ is not a single valued function. Thus,
the resulting expression $B$ in eq. (\ref{eq70}) is not  well
behaved if $\dot\Omega \neq 0$. Consequently, in order to stay
with a physically meaningful result, we must impose the constraint
\begin{equation}
\label{eq71}
\dot\Omega = 0 \,.
\end{equation}
Without any loss of generality, we can also take
\begin{equation}
\label{eqq71}
\Omega = 1 \,.
\end{equation}
The associated Noether symmetries now comprises a time--independent
rotation and two time--dependent spatial translations. The
associated magnetic field, according to the solution (\ref{eq70})
and the restriction (\ref{eq71}), has the form
\begin{equation}
\label{eq72}
B = \bar{B}(\bar{x}, \bar{y}) \,.
\end{equation}
Putting this functional form $B$ into the system
(\ref{eq28}--\ref{eq29}), taking into account $\rho = 0$, $\Omega
= 1$ and the definition (\ref{eqq44}) of $\beta_1$ and $\beta_2$,
yields
\begin{eqnarray}
\label{eq73}
E_{1{\bar{t}}} &=& \hphantom{+} \ddot\beta_2 - E_2 +
\dot{\beta_1}\bar{B}(\bar{x},\bar{y})  \,,\\
\label{eq74}
E_{2{\bar{t}}} &=& - \ddot\beta_1 + E_1 +
\dot{\beta}_{2}\bar{B}(\bar{x},\bar{y}) \,.
\end{eqnarray}
In these equations, $\bar{x}$ and $\bar{y}$ are formally parameters
independent of $\bar{t}$. Consequently, it is not difficult to
obtain the solution, which reads
\begin{eqnarray}
E_1 &=& \ddot\beta_1 -
\dot{\beta}_{2}\bar{B}(\bar{x},\bar{y}) + \nonumber \\
\label{eq75}
&+& \tilde{E}_{1}(\bar{x},\bar{y})\cos\bar{t} -
\tilde{E}_{2}(\bar{x},\bar{y})\sin\bar{t}  \,,\\
E_2 &=& \ddot\beta_2 + \dot{\beta}_{1}\bar{B}(\bar{x},\bar{y}) + \nonumber \\
\label{eq76}
&+& \tilde{E}_{2}(\bar{x},\bar{y})\cos\bar{t} +
\tilde{E}_{1}(\bar{x},\bar{y})\sin\bar{t}  \,,
\end{eqnarray}
where $\tilde{E}_1$ and $\tilde{E}_2$ are arbitrary functions of
$\bar{x}$ and $\bar{y}$.

By defining new arbitrary functions
\begin{equation}
\label{eq79}
\bar{E}_1 = \tilde{E}_{1}/\bar{x} \quad ,
\quad \bar{E}_2 = \tilde{E}_{2}/\bar{x} \quad ,
\end{equation}
we obtain, in physical coordinates,
\begin{eqnarray}
E_1 &=& \ddot\beta_1 - \dot\beta_{2}\bar{B}(\bar{x},\bar{y}) +
\nonumber \\
\label{eq77} &+& (x - \beta_{1})\bar{E}_{1}(\bar{x},\bar{y})
- (y - \beta_{2})\bar{E}_{2}(\bar{x},\bar{y}) \,,
\\
E_2 &=& \ddot\beta_2 + \dot{\beta}_{1}\bar{B}(\bar{x},\bar{y}) +
\nonumber \\
\label{eq78}
&+& (x - \beta_{1})\bar{E}_{2}(\bar{x},\bar{y})
+ (y - \beta_{2})\bar{E}_{1}(\bar{x},\bar{y}) \,,
\end{eqnarray}
where $\bar{x}$, $\bar{y}$ are given in equations
(\ref{eq42}--\ref{eq43}).

Faraday's law now imposes
\begin{equation}
\label{eq80}
\bar{x}\bar{E}_{2\bar{x}} + 2\bar{E}_2 = - \bar{B}_{\bar{y}} \,,
\end{equation}
whose solution is
\begin{equation}
\label{eq81} \bar{E}_2 =
\frac{1}{\bar{x}^2}\frac{\partial\psi}{\partial\bar{y}} \quad ,
\quad \bar{B} = -
\frac{1}{\bar{x}}\frac{\partial\psi}{\partial\bar{x}} \quad ,
\end{equation}
where $\psi = \psi(\bar{x},\bar{y})$ is an arbitrary function.

In conclusion, the electromagnetic field is given by eqs. (\ref{eq72})
and (\ref{eq77}--\ref{eq78}), with the constraint (\ref{eq81}). There
remain four arbitrary functions, namely $E_{1}(\bar{x},\bar{y})$,
$\psi(\bar{x},\bar{y})$, $\beta_{1}(t)$ and $\beta_{2}(t)$, with $\bar{x}$,
$\bar{y}$ defined in equations (\ref{eq42}--\ref{eq43}).

\subsection{The case $\rho = 0$, $\Omega = 0$ and $a_{2} \neq 0$}

The procedure to find the electromagnetic field is by now clear.
We simply list the results. The equation for the magnetic field is
\begin{equation}
\label{eq82}
B_{\bar{t}} = 0 \,,
\end{equation}
or
\begin{equation}
\label{eq83}
B = \bar{B}(\bar{x}, \bar{y}) \,,
\end{equation}
with $\bar{B} = \bar{B}(\bar{x},\bar{y})$ an arbitrary function of
$\bar{x}$ and $\bar{y}$, which are defined by equations
(\ref{eq46}--\ref{eq47}).

The equations for $E_1$ and $E_2$ are
\begin{eqnarray}
\label{eq84} E_{1\bar{t}} &=& \ddot{a}_{1}(\bar{y}) -
\dot{a}_{2}(\bar{y})\bar{B} \,,\\ \label{eq85} E_{2\bar{t}} &=&
\ddot{a}_{2}(\bar{y}) + \dot{a}_{1}(\bar{y})\bar{B} \,,
\end{eqnarray}
with solution
\begin{eqnarray}
\label{eq86} E_1 &=& (\ddot{a}_{1}(\bar{y}) -
\dot{a}_{2}(\bar{y})\bar{B}(\bar{x},\bar{y}))\bar{t} +
\bar{E}_{1}(\bar{x}, \bar{y}) \,,\\ \label{eq87} E_2 &=&
(\ddot{a}_{2}(\bar{y}) +
\dot{a}_{1}(\bar{y})\bar{B}(\bar{x},\bar{y}))\bar{t} +
\bar{E}_{2}(\bar{x}, \bar{y}) \,,
\end{eqnarray}
where $\bar{E}_1$ and $\bar{E}_2$ are arbitrary functions of the
indicated arguments. In physical coordinates,
\begin{eqnarray}
\label{eqq86}
E_1 &=& \frac{\ddot{a}_{1}y}{a_2} -
\frac{\dot{a}_{2}y}{a_2}\bar{B}(\bar{x},\bar{y}) +
\bar{E}_{1}(\bar{x}, \bar{y}) \,,\\
\label{eqq87}
E_2 &=& \frac{\ddot{a}_{2}y}{a_2} +
\frac{\dot{a}_{1}y}{a_2}\bar{B}(\bar{x},\bar{y}) +
\bar{E}_{2}(\bar{x}, \bar{y}) \,,
\end{eqnarray}
where $\bar{x}$, $\bar{y}$ defined in equations
(\ref{eq45}--\ref{eq46}).

After solving the differential equations that arise from
Noether's symmetry condition, we must verify what is the
constraint imposed by Faraday's law. After some calculation using
eqs. (\ref{eq83}), (\ref{eqq86}--\ref{eqq87}) and the form
(\ref{eq45}--\ref{eq47}) of the canonical group coordinates, we
conclude that Faraday's law implies
\begin{equation}
\label{eq88}
\bar{B}_{\bar{y}} + \left(\frac{1}{a_2}(a_{1}\bar{E}_1 +
a_{2}\bar{E}_2 - \ddot{a}_{1}\bar{x} +
\dot{a}_{2}\int^{\bar{x}}B(\mu,\bar{y})d\mu)\right)_{\bar{x}} = 0 \,.
\end{equation}
The solution, in terms of an arbitrary function
$\psi = \psi(\bar{x},\bar{y})$, is
\begin{eqnarray}
\label{eq89} \strut   \bar{B} =
\frac{\partial\psi}{\partial\bar{x}} \,, \qquad \qquad \qquad
\qquad \qquad \qquad \strut \\ \label{eq90}
\frac{1}{a_2}(a_{1}\bar{E}_1 + a_{2}\bar{E}_2 -
\ddot{a}_{1}\bar{x} +
\dot{a}_{2}\int^{\bar{x}}\bar{B}(\mu,\bar{y})d\mu)
=  - \frac{\partial\psi}{\partial\bar{y}} \,,
\end{eqnarray}
where the last equation can be rewritten in terms of a new
function $\bar{V} = \bar{V}(\bar{x}, \bar{y})$ according to
\begin{eqnarray}
\label{eq91}
\bar{E}_1 &=& -
\bar{V}_{\bar{x}} \,,\\
\label{eq92}
\bar{E}_2 &=& \frac{\ddot{a}_1}{a_2}\bar{x}
- \frac{\dot{a}_2}{a_2}\psi
- \psi_{\bar{y}}
+ \frac{a_1}{a_2}\bar{V}_{\bar{x}} \,.
\end{eqnarray}

This completely determines the last class of solutions for the
electromagnetic field. $B$ is given by eq. (\ref{eq83}) and $E_1$
and $E_2$ are given by eqs. (\ref{eqq86}--\ref{eqq87}). The
functions $\bar{B}$, $\bar{E}_1$ and $\bar{E}_2$, present in the
solution, are given by eqs. (\ref{eq89}) and
(\ref{eq91}--\ref{eq92}), in terms of the arbitrary functions
$\psi(\bar{x}, \bar{y})$ and $\bar{V}(\bar{x}, \bar{y})$
with $\bar{x}$, $\bar{y}$ defined in equations (\ref{eq45}--\ref{eq46}).
The arbitrary functions $a_{1}(t)$ and $a_{2}(t)$ are also present in
the electromagnetic field, so that four arbitrary functions
participate in the final solution.

\section{Conserved quantities}

The electromagnetic potentials, which are gauge dependent, are a
basic ingredient in the Noether's invariant (\ref{eqq9}).
However, the resulting Noether's constant of motion is
always independent of gauge choice, as seen in the continuation.

To find the Noether constant of motion corresponding
to each of the point symmetry, we must also solve equations
(\ref{eq24}--\ref{eq26}) for the function $F(x,y,t)$ which appears
in the definition (\ref{eqq9}). The system
(\ref{eq24}--\ref{eq26}) is solvable for $F$ by construction,
since the electromagnetic fields derived in the last section
satisfy the basic equations (\ref{eq27}--\ref{eq29}). These
equations, in turn, are the necessary and sufficient conditions
for the existence of a solution $F$, satisfying system
(\ref{eq24}--\ref{eq26}).

Given the electromagnetic fields and the related symmetries, it is not
difficult to find the appropriate electromagnetic potentials and the
associated function $F$. We only show the results pertaining to
each type of solution.

\subsection{The case $\rho \neq 0$}

The vector potential for the magnetic field listed in (\ref{eq49}) is
\begin{equation}
\label{eq93} {\bf A} = \frac{{\bf q}\times{\bf\Omega}}{\rho^2} +
\frac{R(T)\cdot{\bf\bar{A}}(\bar{x},\bar{y})}{\rho} + \nabla\lambda \,,
\end{equation}
for arbitrary gauge function $\lambda =
\lambda(x,y,t)$. The gauge
function $\lambda$ is irrelevant in the calculation of the
magnetic field. However, it was kept in order to show
explicitly the gauge independence of Noether's constant of motion.
Furthermore, ${\bf\bar{A}} = (\bar{A}_{1}(\bar{x},\bar{y}),\,
\bar{A}_{2}(\bar{x},\, \bar{y}),\, 0)$ is a vector satisfying
\begin{equation}
\label{eq94} \bar{A}_{2\bar{x}} - \bar{A}_{1\bar{y}} = \bar{B} \,,
\end{equation}
where $\bar{B}$ is defined in eq. (\ref{eq49}).

The scalar potential is
\begin{eqnarray}
V &=& - (\rho\ddot{\bf\alpha} - \ddot\rho{\bf\alpha})\cdot\frac{\bf q}{\rho} -
\frac{\ddot\rho}{2\rho}{\bf q}^2 - \frac{1}{\rho^3}(\rho\dot{\bf\alpha} -
\dot\rho{\bf\alpha})\times{\bf\Omega}\cdot{\bf q} -
\frac{1}{2\rho^4}({\bf\Omega}\times{\bf q})^2 + \nonumber
\\ \label{eq95} &+& \frac{1}{\rho^2}\bar{V}(\bar{x},\bar{y}) +
\frac{{\bf\eta}\cdot\, R(T)\cdot{\bf \bar{A}}(\bar{x},\bar{y})}{\rho^3} - \lambda_t
\,.
\end{eqnarray}

The function $F$ is calculated using the electromagnetic potentials
and eqs. (\ref{eq24}--\ref{eq26}). The result is
\begin{eqnarray}
F &=& \frac{1}{2}(\rho\dot{{\bf\alpha}} - \dot\rho{\bf\alpha})^2 +
\dot\rho(\rho\dot{{\bf\alpha}} - \dot\rho{\bf\alpha})\cdot{\bf q} +
\rho(\rho\ddot{{\bf\alpha}} - \ddot\rho{\bf\alpha})\cdot{\bf q} + \nonumber \\
\label{eq96} &+& \frac{1}{2}(\dot\rho^2 + \rho\ddot\rho){\bf q}^2 +
\frac{1}{\rho}(\rho\dot{{\bf\alpha}} - \dot\rho{\bf\alpha})\cdot{\bf\Omega}\times{\bf
q} + G\,\lambda \,.
\end{eqnarray}
Notice the presence of the gauge function $\lambda$ in the last term of
the equation (\ref{eq96}).

All the ingredients to construct the Noether invariant
(\ref{eqq9}) are already obtained. We arrive at
\begin{equation}
\label{eq97} I = \frac{1}{2}\left(\rho(\dot{\bf q} - \dot{{\bf\alpha}}) -
\dot\rho({\bf q} - {\bf\alpha}) - \frac{{\bf\Omega}\times{\bf q}}{\rho} \right)^2 +
\bar{V}(\bar{x},\bar{y}) \,.
\end{equation}

Remarks:

a) For $\Omega = 0$, the invariant (\ref{eq97}) recovers the
two--dimensional version of an invariant derived by Lewis
\cite{Lewis}, in his search for quadratic invariants for
three--dimensional non--relativistic charged particle motion. For
$\Omega \neq 0$, that is, when we include also rotations, our
result is new.

b) Interestingly, despite the dependence of
the electromagnetic potentials and the
function $F$ on the gauge function $\lambda$, the resulting
invariant (\ref{eq97}) is gauge independent. This is just what we
expected a priori, since the choice of gauge should have no influence on
physical quantities.

To provide an interpretation of the invariant, we make a change
of variables. Using canonical group coordinates $\bar{\bf q} =
(\bar{x},\bar{y},0)$ and $\bar{t}$, the Lagrangian
function reads
\begin{equation}
\label{eqq99}
L = \frac{1}{\rho^2}\bar{L}(\bar{\bf q}', \bar{\bf q}) + \frac{dW}{dt} \,,
\end{equation}
where
\begin{eqnarray}
\label{eqq100} \bar{L}(\bar{\bf q}', \bar{\bf q}) &=& \frac{1}{2}\bar{\bf q}'^2 +
\bar{\bf A}(\bar{\bf q})\cdot\bar{\bf q}' - \bar{V}(\bar{\bf q}) \,,\\ W &=& \lambda +
\frac{\dot\rho{\bf q}^2}{2\rho} + (\rho\dot{{\bf\alpha}} -
\dot\rho{\bf\alpha})\cdot\frac{\bf q}{\rho} + \nonumber
\\ \label{eqq101} &-&
\int^{t}\frac{d\mu}{\rho^{2}(\mu)} (\rho(\mu)\dot{{\bf\alpha}}(\mu) -
\dot\rho(\mu){\bf\alpha}(\mu))^2 \,.
\end{eqnarray}
The primes denote differentiation with respect to $\bar{t}$. The function $W$ defined
by eq. (\ref{eqq101}) can be disregarded in the Lagrangian, as it only adds a total
derivative. Moreover, using $\bar{t}$ as the new time parameter, the action functional
(\ref{eq0}) reads
\begin{equation}
\label{eqq102}
S = \int\,\bar{L}(\bar{\bf q}', \bar{\bf q})d\bar{t} \,,
\end{equation}
from which it is evident that $\bar{L}$ may be used as the Lagrangian
for the motion described in terms of canonical group coordinates.
Indeed, using $\bar{L}$, the Lorentz equations become
\begin{eqnarray}
\label{eq98}
\bar{x}'' &=& - \bar{V}_{\bar{x}}(\bar{x},\bar{y}) +
\bar{y}'\,\bar{B}(\bar{x},\bar{y}) \,,\\
\label{eq99}
\bar{y}'' &=& - \bar{V}_{\bar{y}}(\bar{x},\bar{y}) -
\bar{x}'\,\bar{B}(\bar{x},\bar{y}) \,,
\end{eqnarray}
which are the equations for two--dimensional non--relativistic
charged particle motion in a time--independent electromagnetic
field. We observe that the relation
\begin{equation}
\label{eq100} \bar{\bf q}' = \frac{d\bar{\bf q}}{d\bar{t}} =
\rho\,R^{-1}(T)\cdot\left(\dot{\bf q} - \frac{\bf\eta}{\rho^2}\right)
\end{equation}
is useful to obtain the transformed Lagrangian and
Lorentz equations.

The transformed equations of motion, being autonomous, have an
associated energy-like invariant, of the form
\begin{equation}
\label{eq101}
I = \frac{1}{2}\bar{\bf q}'^2 + \bar{V}(\bar{x},\bar{y}) \,,
\end{equation}
which is precisely the Noether invariant (\ref{eq97}) in transformed
coordinates. Thus, the Noether constant of motion is the energy
expressed in the variables where the equations of motion are
autonomous.

\subsection{The case $\rho = 0$ and $\Omega \neq 0$}

The vector potential is, in this case,
\begin{equation}
\label{eq102}
{\bf A} = (y - \beta_{2}, - (x - \beta_{1}),
0)\frac{\psi(\bar{x},\bar{y})}{\bar{x}^2} + \nabla\lambda \,,
\end{equation}
where $\lambda = \lambda(x,y,t)$ is the gauge function.
The scalar potential $V$ is now given by
\begin{eqnarray}
V &=& - \ddot\beta_{1}(x - \beta_{1}) -
\ddot\beta_{2}(y - \beta_{2}) + \bar{V}(\bar{x},\bar{y}) +
\nonumber \\
\label{eq103}
&+& \frac{1}{\bar{x}^2}\left(\dot{\beta}_{1}(y - \beta_{2}) -
\dot{\beta}_{2}(x - \beta_{1})\right)\psi(\bar{x},\bar{y})
- \lambda_t \,,
\end{eqnarray}
where
\begin{equation}
\label{eq104}
\bar{V}_{\bar{x}} = - \bar{x}\bar{E}_{1}(\bar{x},\bar{y}) \,.
\end{equation}
The resulting function $F$ is
\begin{equation}
\label{eq105} F = \dot{\beta}_{2}(x - \beta_{1}) -
\dot{\beta}_{1}(y - \beta_{2}) + G\lambda \,,
\end{equation}
and the associated Noether invariant is
\begin{equation}
\label{eq106}
I = (y - \beta_{2})(\dot x - \dot\beta_{1}) -
(\dot{x} - \beta_{1})(y - \beta_{2}) + \psi(\bar{x},\bar{y})   \,.
\end{equation}

To interpret Noether's invariant, again we change variables. Let
the new coordinates of configuration space be
$(\bar{x}, \bar{t})$. Using these variables, the Lagrangian becomes
\begin{equation}
\label{eq107}
L = \bar{L} + \frac{dW}{d\bar{y}} \,,
\end{equation}
where
\begin{equation}
\label{eq108}
\bar{L} = \frac{1}{2}(\dot{\bar{x}}^2 +
\bar{x}^{2}\dot{\bar{t}}^{2})
- \dot{\bar{t}}\psi(\bar{x},\bar{y}) - \bar{V}(\bar{x},\bar{y})
\end{equation}
is a new Lagrangian and
\begin{equation}
\label{eq109}
W = \lambda + (\dot{\beta}_{1}\cos\bar{t} +
\dot{\beta}_{2}\sin\bar{t})\bar{x} +
\frac{1}{2}\int^{t}\left(\dot{\beta}_{1}^{2}(\mu) +
\dot{\beta}_{2}^{2}(\mu)\right)d\mu \,.
\end{equation}
In this new description, $(\bar{x},\bar{t})$ are the dependent variables
and $\bar{y}$ is the independent variable.

In equation (\ref{eq107}), it is apparent that the Lagrangian in
the new coordinates can be taken as simply $\bar{L}$, since the
addition of a total time derivative does not influence the
equations of motion. Since $\bar{t}$ is a cyclic coordinate the momentum
conjugate to $\bar{t}$,
\begin{equation}
\label{eq110}
p_{\bar{t}} = \bar{L}_{\dot{\bar{t}}} = \bar{x}^{2}\dot{\bar{t}}
- \psi(\bar{x},\bar{y}) \,,
\end{equation}
is a conserved quantity. This conserved quantity (\ref{eq110}) is, apart
from an irrelevant sign, the Noether invariant (\ref{eq106}), which
we can be interpreted as the
conserved momentum conjugated to
the cyclic coordinate $\bar{t}$.

\subsection{The case $\rho = 0$, $\Omega = 0$ and $a_{2} \neq 0$}

Now, the vector potential is
\begin{equation}
\label{eq112}
{\bf A} = (0, \psi, 0) + \nabla\lambda \,.
\end{equation}
where $\lambda = \lambda(x,y,t)$ is the gauge function, while the
scalar potential is
\begin{equation}
\label{eq113}
V = - \frac{\ddot{a}_1}{a_2}xy + \frac{1}{2a_{2}^2}(a_{1}\ddot{a}_1 -
a_{2}\ddot{a}_{2})\,y^2 + \bar{V}(\bar{x},\bar{y}) +
\frac{\dot{a}_{2}y}{a_2}\psi(\bar{x},\bar{y}) - \lambda_t \,.
\end{equation}
Using these electromagnetic potentials, we arrive at the
function
\begin{equation}
\label{eq114} F = \dot{a}_{1}\,x + \dot{a}_{2}\,y + G\lambda \,,
\end{equation}
so that the Noether invariant is
\begin{equation}
\label{eq115} I = - (a_{1}\dot{x} + a_{2}\dot{y} - \dot{a}_{1}x -
\dot{a}_{2}y + a_{2}\psi(\bar{x},\bar{y})) \,.
\end{equation}
Again Noether's invariant is a gauge independent quantity.

For the interpretation of the invariant, we use $\bar{x}$ and
$\bar{t}$ as new dependent variables and $\bar{y}$ as new
independent variable. The Lagrangian can be expressed as
\begin{equation}
\label{eq116}
L = \bar{L} + \frac{dW}{d\bar{y}} \,,
\end{equation}
where now
\begin{equation}
\label{eq117}
\bar{L} = \frac{1}{2}(\dot{\bar{x}}^2 +
(a_{1}^2 + a_{2}^2)\dot{\bar{t}}^2) +
(a_{1}\dot{\bar{x}} - \dot{a}_{1}\bar{x})\dot{\bar{t}} + a_{2}\psi(\bar{x},
\bar{y})\dot{\bar{t}} - \bar{V}(\bar{x}, \bar{y})
\end{equation}
is a new Lagrangian function, and,
\begin{equation}
\label{eq118}
W = \lambda +
\dot{a}_{1}\bar{x}\bar{t} + \frac{1}{2}(a_{1}\dot{a}_{1} +
a_{2}\dot{a}_{2})\bar{t}^2 \,.
\end{equation}
The function $W$ can be disregarded in the Lagrangian, as it
enters $\bar{L}$ merely in the form of a total time derivative. For the
Lagrangian $\bar{L}$, $\bar{t}$ is a cyclic
variable, yielding the conserved momentum
\begin{equation}
\label{eq119}
p_{\bar{t}} = \bar{L}_{\dot{\bar{t}}} = (a_{1}^{2} + a_{2}^{2})\dot{\bar{t}}
+ a_{1}\dot{\bar{x}} - \dot{a}_{1}\bar{x} + a_{2}\psi(\bar{x},\bar{y}) \,,
\end{equation}
which, apart from an irrelevant sign, is the Noether invariant
(\ref{eq115}). In conclusion, the Noether invariant may be interpreted
as the momentum conjugated to the cyclic coordinate $\bar{t}$.

\section{Conclusion}

We have found the class of electromagnetic fields compatible with
Noether symmetries for Lagrangians of type (\ref{eq11}),
describing physically interesting non--relativistic charged
particle motion. The treatment comprises the complete resolution
of the basic system of partial differential equations
(\ref{eq27}--\ref{eq29}) by use of canonical group coordinates.
There are three classes of electromagnetic fields yielding action
functionals endowed with Noether invariance, as listed in section
4. These electromagnetic fields are consistent with Maxwell
equations and depend on several arbitrary functions. The
corresponding Noether invariants were explicitly shown in section
5, one in the form of an energy-like function (\ref{eq97}) and two
in the form of momentum-like functions, (\ref{eq106}) and
(\ref{eq115}).

For a possible extension of the present work, we mention the
investigation of the fully three--dimensional case. While trivial
in principle, this extension may present, in practice, difficult
mathematical problems. As a further development, one can analyze
the complete integrability of the Lorentz equation corresponding
to the electromagnetic fields associated with Noether point
symmetries. From Liouville's theorem, two constants of motion are
sufficient for the complete integrability of the equations of
motion of a Hamiltonian system with two degrees of freedom, as in
the present case. The explicit dependence on time does not modify
this statement \cite{Bouq}. In the present work, we have derived
classes of electromagnetic fields yielding just one constant of
motion. For complete integrability, there is the need for a second
invariant. This quantity  exist only for special forms, found
within the classes of electromagnetic fields constructed here.
Consequently, the complete integrability of these systems remains
an open question. As a final remark, the Noether invariants
derived here may be useful in the construction of exact
time--dependent solutions for the self-consistent Vlasov-Maxwell
system in plasma physics.

\bigskip\noindent
{\bf{\large Acknowledgement}}\\
This work has been partially supported by Funda\c{c}\~ao de Amparo a Pesquisa do
Estado do Rio Grande do Sul (FAPERGS). We also gratefully acknowledge one of the
referees for valuable comments and suggestions.

\end{document}